# Two novel pure-state coherence measures in quantifying coherence

Manis Hazra[1] and Debabrata Goswami[2]
*Indian Institute of Technology Kanpur, Kanpur-208016, India*

In the resource theory of coherence, the quantification of quantum-state coherence is an important task. In this regard, the key ingredients are the various coherence monotones (or measures). There are few coherence-monotone classes that solely depend on other coherence measures defined for all the pure states; in other words, they rely on the pure-state coherence measures (PSCM). Here, we set forth two such novel PSCMs, and validate each of them through the fulfillment of all four necessary conditions. In addition, we delve into the most recent (as per our knowledge) coherence-monotone class based on the innovative idea of quantifying coherence in terms of pure-state coherence, further redefine it, and, through the study of convexity under mixing, justify why this coherence-monotone class cannot be treated as a coherence-measure class in general.

## I. INTRODUCTION

Coherence, as depicted by the superposition among the quantum levels in a fixed reference frame, is a fundamental property of quantum theory. It emphasizes most of the quantum features, such as quantum correlation [1, 2], entanglement [3–5], and symmetry [6]. It also accounts for quantum advantages in many fields, including quantum thermodynamics [7–11], quantum metrology [12–14], quantum cryptography [15, 16], quantum biology [17, 18], etc. The recent development of the resource theory of coherence [19, 20] endows a robust mathematical framework for quantifying coherence [21, 22], along with establishing a platform for intense analysis of the effect of coherence in fundamental physics [23].

In the resource theory of coherence [19, 21, 24–27], a major portion is dedicated to the quantification of coherence. In this regard, many functionals within the coherence framework have been proposed so far, and its search is continuing. Based on methodology, the measure of the lowest distance [21, 28–30] of the state of interest from the set of free (or incoherent) states (which is unambiguously defined) may be the most intuitive idea. The $l_1$-norm of coherence [21], the relative entropy of coherence [21], and geometric coherence [3] are impressive examples in this category. Apart from distance-based measures, the convex-roof technique [31], quantification based on entanglement monotone [3], and quantification in terms of pure-state coherence [32] provide unique mechanisms that churn out different classes of *coherence monotones* (or *measures*). The finding of novel coherence measures is always advantageous as it adds new computability as well as broadens the operational aspect, thus significantly contributing toward enhancing resource theory [33–36].

In the mixed state domain, apart from the traditional class, i.e., convex-roof construction ($C_{conv.roof}$), the latest one (as per our knowledge) is the quantification of coherence in terms of pure-state coherence ($C_P$). $C_P$ relies on the innovative idea where it drives for the pure state of minimal coherence from which the mixed state of interest can be achieved via incoherent operation (IO) [32]. One common thing between $C_{conv.roof}$ and $C_P$ is that both these *classes* depend on the concave, symmetric functions, called the *pure-state coherence measure* (PSCM) [31, 32]. Hereafter, we will use the mathematical notation $\check{C}$ to represent any PSCM. As the name suggests, $\check{C}$ are defined for all the pure states. the classes $C_{conv.roof}$ and $C_P$ creäte unique *coherence monotones* (*measures*) when different $\check{C}$ are put into operation [31, 32]. Therefore, a bona fide $\check{\mathcal{C}}$ is considered critical in achieving distinctive *coherence measures*, or *monotones*. In this paper, we put forth two such *pure-state coherence measures* ($\check{\mathcal{C}}$). In addition, we try to contribute towards the insight of $\mathcal{C}_P$ through redefinition (of $\mathcal{C}_P$) and comment regarding its' convexity under mixing.

---
[1] manisin.123@gmail.com
[2] dgoswami@iitk.ac.in

The rest of the paper is set up as follows: In Sec. II, we briefly review the basics of the framework of the coherence resource theory (especially the coherence criteria), then discuss the methodology of $\mathcal{C}_P$. In Sec. III, we first provide the four necessary conditions for any bona fide PSCM and thereafter set forth two new PSCMs and prove their validity based on these conditions. We further define their *coherence-monotone* (*measure*) counterparts depending on $\mathcal{C}_P$ and $\mathcal{C}_{conv.roof}$. In Sec. IV, we redefine $\mathcal{C}_P$, along with provide the reason why $\mathcal{C}_P$ cannot be considered a *coherence-measure* class. Lastly, in Sec. V, with some necessary remarks, we draw the conclusion.

## II. THEORETICAL BACKGROUND

### A. Resource Theory of coherence Framework

As with the resource theory of quantum entanglement or any other quantum resource [20, 33], the framework of coherence resource theory is well defined by free operations and free states. In the reference basis, $\{|i\rangle\}$, a free or incoherent state is denoted by $\delta$, where $\delta = \sum_i \delta_i |i\rangle\langle i|$ with the basis-state probabilities $\delta_i$. Thus, the incoherent class $\mathfrak{T}: \delta \in \mathfrak{T}$ contains the states abstaining from any non-zero off-diagonal elements in their density-matrix forms. The free (or incoherent) operation ($\Lambda$) is a completely positive and trace-preserving (CPTP) map that cannot increase coherence. Therefore, it maps an incoherent state to any other incoherent state: $\Lambda\delta\Lambda^\dagger \in \mathfrak{T}$. $\Lambda$ is defined in the Kraus representation as $\Lambda(\cdot) = \sum_n K_n(\cdot)K_n^\dagger$ with $K_n(\cdot)K_n^\dagger \in \mathfrak{T}$ and $\sum_n K_n K_n^\dagger = I$, where $I$ denotes the unit matrix and "·" stands for an arbitrary quantum state. The coherence framework sets the following criteria that a bona fide coherence quantifier ($\mathcal{C}$) should fulfill [21, 32]:

C1: *Non-negativity*. For any quantum state $\rho$, $\mathcal{C}(\rho) = 0 \, \forall \, \rho \in \mathfrak{T}$, otherwise $\mathcal{C}(\rho) > 0$.

C2: *Monotonicity*. $\mathcal{C}(\rho) \geq \mathcal{C}(\Lambda\rho\Lambda^\dagger)$. This criterion basically reflects the definition of free or incoherent operation, $\Lambda$.

C3: *Strong monotonicity*. state coherence should not increase under sub-selection measurements on average, i.e., $\mathcal{C}(\rho) \geq \sum_n p_n \mathcal{C}(K_n \rho K_n^\dagger / p_n) = \sum_n p_n \mathcal{C}(K_n \rho_n K_n^\dagger)$, with $p_n = Tr[K_n \rho K_n^\dagger]$, the probability that the incoherent Kraus operation $K_n$ acting on $\rho$ (in other words, $p_n$ is expectation value of the state $\rho_n = K_n \rho K_n^\dagger / Tr[K_n \rho K_n^\dagger]$, achieved after the action of $K_n$).

C4: *Convexity*. for any mixed state $\rho \, (= \sum p_i \rho_i)$, the coherence of $\rho$ is not greater than the average coherence of the participating pure states $\rho_i$, i.e., $\mathcal{C}(\rho) \leq \sum p_i \mathcal{C}(\rho_i)$.

C5: *Maximal coherence*. $\mathcal{C}(\rho) < \mathcal{C}(|\Phi_d\rangle\langle\Phi_d|)$, for any $\rho$ other than $|\Phi_d\rangle\langle\Phi_d|$, where $|\Phi_d\rangle (= \frac{1}{\sqrt{d}}\sum_{n=1}^d e^{i\theta_n}|n\rangle$ with real $\theta_n$) [37] is the set of *maximally coherent states* (MCS) of dimension $d$.

Note: When $\mathcal{C}(\rho)$ satisfies all five criteria C1–C5, it is classified as a *coherence measure*; conversely, if it satisfies all the given criteria except C4, it is called a *coherence monotone*, in the same way as an entanglement monotone. Due to its' *convex* (C4) nature, the handling of a *coherence measure* is easier than a *coherence monotone* in terms of mathematical convenience [32]. However, it is seen in some instances that *coherence monotones* have an operational advantage over *coherence measures* and thus play an important role as coherence quantifiers [38].

### B. Concept of $\mathcal{C}_P$

In quantum state transformation, an objective state $\rho$ can be achieved from another state $\varphi$ through incoherent operations (IO), but only if the coherence of $\varphi$ is not less than $\rho$ [21], i.e., $\varphi \xrightarrow{IO} \rho$ iff $\mathcal{C}(\varphi) \geq \mathcal{C}(\rho)$ for $\mathcal{C}$ being any valid coherence monotone or measure. The above statement implies that a general state $\rho$ (pure, or mixed) can be realized from different pure states through distinct incoherent channels. In this sense, *for a particular $\rho$, the set of pure states is non-empty*. This non-empty set for $\rho$ is denoted by $R(\rho)$. In ref. [32], it is shown that the minimal coherence achieved from the set of pure states $|\varphi\rangle \in R(\rho)$ can be a valid coherence quantifier of $\rho$.

***Theorem 1*** [32]. If $\rho$ is a given state (pure or mixed) that can be converted from a set of pure states $|\varphi\rangle$ in $R(\rho)$ through IO, then $\mathcal{C}_P(\rho)$ is a coherence monotone with

$$\mathcal{C}_P(\rho) = \inf_{|\varphi\rangle \in R(\rho)} \check{\mathcal{C}}(|\varphi\rangle). \quad (1)$$

Here, $\check{\mathcal{C}}(\cdot)$ is any *pure-state coherence measure (PSCM)*. All the bona fide coherence monotones can be used in Eq. (1) as $\check{\mathcal{C}}(\cdot)$; the reason is that a valid coherence monotone must be defined for all the pure states. Note that a PSCM should fulfill a set of conditions, discussed in the next section.

Let $|\psi\rangle \in \mathcal{H}^d$ be an arbitrary pure state in a $d$-dimensional Hilbert space ($\mathcal{H}^d$). Then, its *coherence vector* [32] $\boldsymbol{\mu}(|\psi\rangle) \equiv (\langle 1|\psi\rangle\langle\psi|1\rangle, \langle 2|\psi\rangle\langle\psi|2\rangle, \ldots, \langle d|\psi\rangle\langle\psi|d\rangle)^T$. $\boldsymbol{\mu}(|\psi\rangle)$ contains all the diagonal elements of the density matrix $|\psi\rangle\langle\psi|$, in a fixed basis representation $\{|i\rangle\}_{i=1}^d$ (as coherence is basis-dependent). It can be shown in simple terms that any PSCM ($\check{\mathcal{C}}$) is a function of the diagonal elements of the density matrix $|\psi\rangle\langle\psi|$ for any pure state $|\psi\rangle$; in other words, $\check{\mathcal{C}}(|\psi\rangle) \equiv \check{\mathcal{C}}(\boldsymbol{\mu}(|\psi\rangle))$. Therefore, to find a new coherence monotone of class $\mathcal{C}_P$, the search for a function of the coherence vector $\boldsymbol{\mu}$ is critical. In Sec. III, we propose two such PSCMs.

## III. TWO NEW PSCMs: PROPOSITIONS AND PROOFS

In the pure-state domain, any coherence measure $\mathcal{C}$ can be effectively reduced to its PSCM counterpart $\check{\mathcal{C}}$, which is a symmetric, concave function of $\boldsymbol{\mu}(\rho)$. A bona fide PSCM must obey the following four conditions [39, 40]:

*Condition 1*. If $\boldsymbol{\mu}$ is any permutation of $(1,0,\ldots,0)^T$, $\check{\mathcal{C}}(\boldsymbol{\mu}) = 0$.

*Condition 2*. Under any permutation operation $P_\pi$, $\check{\mathcal{C}}(\boldsymbol{\mu})$ is invariant, i.e., $\check{\mathcal{C}}(P_\pi(\boldsymbol{\mu})) = \check{\mathcal{C}}(\boldsymbol{\mu})$; here, $P_\pi$ is the permutation matrix corresponding to $\pi$, which is a permutation of $\{1,2,\ldots,d\}$.

*Condition 3*. $\check{\mathcal{C}}(\boldsymbol{\mu})$ should be a concave function of $\boldsymbol{\mu}$: $\check{\mathcal{C}}(\lambda\boldsymbol{\mu}(\rho_1) + (1-\lambda)\boldsymbol{\mu}(\rho_2)) \geq \lambda\check{\mathcal{C}}(\boldsymbol{\mu}(\rho_1)) + (1-\lambda)\check{\mathcal{C}}(\boldsymbol{\mu}(\rho_2))$.

*Condition 4*. $\check{\mathcal{C}}(\boldsymbol{\mu})$ reaches the maximal ($\check{\mathcal{C}}(\boldsymbol{\mu}) = 1$, in the case of normalized PSCM) only if all the elements of $\boldsymbol{\mu}$ is $\frac{1}{d}$, i.e., $\boldsymbol{\mu} = \left(\frac{1}{d}, \frac{1}{d}, \ldots, \frac{1}{d}\right)^T$.

For any incoherent pure state, $\boldsymbol{\mu}$ always takes the form of any of the permutations of $(1,0,\ldots,0)^T$. Therefore, *condition-1* basically signifies the coherence measure criterion, C1 (see Sec. II-A). *Condition-2* arises from the symmetric nature of $\check{\mathcal{C}}(\boldsymbol{\mu})$. Again, concavity fulfills *strict monotonicity*, C3 (see *Theorem1* of Ref. [39]); hence, $\check{\mathcal{C}}(\boldsymbol{\mu})$ must be a concave function (i.e., *condition-3*). Lastly, *condition-4* directly resembles the criterion C5, indicating the *maximally coherent states* (MCS). Keeping in mind all these simple requirements we put forth two new PSCMs in the following that can be converted to coherence monotones or measures in the general scenario by applying them in the classes $\mathcal{C}_P$ and $\mathcal{C}_{conv.roof}$.

### A. Diagonal Difference of PSCM ($\check{\mathcal{C}}_{DD}$)

***Proposition 1:*** Let $\rho$ be any pure-state density matrix in $\mathcal{H}^d$ with its coherence vector $\boldsymbol{\mu}(\rho) = (\rho_{11}, \rho_{22}, \ldots, \rho_{dd})^T$, where $\{\rho_{ii}\}_{i=1}^d$ are the diagonal entries of $\rho$. Then the following function $\check{\mathcal{C}}_{DD}(\rho)$ containing the *modulus* of the difference between any two different elements of $\boldsymbol{\mu}(\rho)$ in all possible combinations, acts as a coherence measure of $\rho$, i.e.,

$$\check{\mathcal{C}}_{DD}(\rho) = (d-1) - \sum_{i=1}^d \sum_{j=i+1}^d |\rho_{ii} - \rho_{jj}|. \quad (2)$$

*Proof.* In the following, we establish $\check{\mathcal{C}}_{DD}(\rho)$ as a valid PSCM through the fulfillment of all four necessary criteria, from *condition-1* to *condition-4*.

*Proof of condition-1.* When the pure state $\rho$ is incoherent, $\boldsymbol{\mu}(\rho)$ is confined to any of the permutation of $(1,0,\ldots,0)^T$. In the RHS of Eq. (2), $\sum_{i=1}^d \sum_{j=i+1}^d |\rho_{ii} - \rho_{jj}|$ is the sum of $C_2^d$ (i.e., $\frac{d(d-1)}{2}$) absolute terms, $|\rho_{ii} - \rho_{jj}|$. For $\boldsymbol{\mu}(\rho) = (1,0,\ldots,0)^T$, we can simply see that $|\rho_{ii} - \rho_{jj}| = 0 \,\forall\, i \neq 1$, whereas $|\rho_{ii} - \rho_{jj}| = 1 \,\forall\, i = 1$; therefore, it gives $\sum_{i=1}^d \sum_{j=i+1}^d |\rho_{ii} - \rho_{jj}| = (d-1)$. Similarly, when $\boldsymbol{\mu}(\rho) = (0,1,...,0)^T$, only $|\rho_{11} - \rho_{22}|$ and $(d-2)$ terms involving $i = 2$ are non-zero (each of them giving one), which again make the sum $(d-1)$. Continuing this way for all possible $\boldsymbol{\mu}(\rho)$, one can deduce that for any pure state $\rho \in \mathfrak{T}$, $\sum_{i=1}^{d}\sum_{j=i+1}^{d}|\rho_{ii} - \rho_{jj}| = (d-1)$. Apply of this result to Eq. (2) shows that $\check{\mathcal{C}}_{DD}(\rho)$ meets *condition-1*.

*Proof of condition-2.* Let $\boldsymbol{\mu}(\rho_0) = (\rho_{11}, \rho_{22}, ..., \rho_{dd})^T$, and after a permutation operation (where the positions of $\rho_{ii}$ and $\rho_{11}$ are interchanged) we have $\boldsymbol{\mu}(\rho_1) = P_\pi(\boldsymbol{\mu}(\rho_0)) = (\rho_{ii}, \rho_{22}, ..., \rho_{11}, ..., \rho_{dd})^T$. As $|\rho_{ii} - \rho_{jj}| = |\rho_{jj} - \rho_{ii}| \forall (i,j)$, it is obvious from Eq. (2) that $\check{\mathcal{C}}_{DD}(\boldsymbol{\mu}(\rho_0)) = \check{\mathcal{C}}_{DD}(\boldsymbol{\mu}(\rho_1))$. Therefore, $\check{\mathcal{C}}_{DD}$ is permutation invariant or a symmetric function of $\boldsymbol{\mu}$.

*Proof of condition-3.* A function $f: \mathbb{R}^n \to \mathbb{R}$ is *concave* if its domain $dom(f)$ is a *convex set*, and for $\forall x, y \in dom(f)$ and $\forall \lambda \in [0,1]$, the following inequality holds [41]:

$$f(\lambda x + (1-\lambda)y) \geq \lambda f(x) + (1-\lambda)f(y) . \quad (3)$$

Conversely, $f(\cdot)$ is *convex* when the inequality is reversed.

In this case, $\boldsymbol{\mu}(\rho) \in \mathbb{R}^d$ always belongs to a *convex set*, as the elements of $\boldsymbol{\mu}(\rho)$ i.e., $\{\rho_{ii}\}$ are constrained by the following two conditions: $\forall \rho_{ii} \geq 0$ and $\sum_{i=1}^{d}\rho_{ii} = 1$. It can be easily verified that $f = \sum_{i=1}^{d}\sum_{j=i+1}^{d}|\rho_{ii} - \rho_{jj}|$ is a *convex* function of $\boldsymbol{\mu}(\rho)$; two illustrations in its support are the following: Let $\boldsymbol{\mu}(\rho_1) = (0.2, 0.3, 0.5)^T$, $\boldsymbol{\mu}(\rho_2) = (0.4, 0.3, 0.3)^T$ and $\lambda = 0.5$; then $f(\lambda \boldsymbol{\mu}(\rho_1) + (1-\lambda)\boldsymbol{\mu}(\rho_2)) = 0.2$, whereas $\lambda f(\boldsymbol{\mu}(\rho_1)) + (1-\lambda)f(\boldsymbol{\mu}(\rho_2)) = 0.4$, thus obeying *convexity*. In another situation, let $\lambda = 0.25$, keeping the *coherence vectors* the same. Here again, the result follows *convexity*: $f(\lambda \boldsymbol{\mu}(\rho_1) + (1-\lambda)\boldsymbol{\mu}(\rho_2)) = 0.1$ and $\lambda f(\boldsymbol{\mu}(\rho_1)) + (1-\lambda)f(\boldsymbol{\mu}(\rho_2)) = 0.3$. Therefore, the presence of a negative sign prior $\sum_{i=1}^{d}\sum_{j=i+1}^{d}|\rho_{ii} - \rho_{jj}|$ in Eq. (2) leaves $\check{\mathcal{C}}_{DD}(\rho)$ concave.

*Proof of condition-4.* All the diagonal elements $\{\rho_{ii}\}$ of any *MCS* (which is obviously a pure state) are of the same value $\frac{1}{d}$, i.e., $\mu(MCS) = \left(\frac{1}{d}, \frac{1}{d}, ..., \frac{1}{d}\right)^T$. Thus, for any MCS, $\sum_{i=1}^{d}\sum_{j=i+1}^{d}|\rho_{ii} - \rho_{jj}| = 0$. Applying this result in Eq. (2), it gives that $\check{\mathcal{C}}_{DD}(MCS) = (d-1)$, which is the maximal (as $\sum_{i=1}^{d}\sum_{j=i+1}^{d}|\rho_{ii} - \rho_{jj}| \geq 0$). So, *condition-4* is satisfied, and with this, the proof of *proposition-1* is completed. ∎

### B. Diagonal Multiplication of PSCM ($\check{\mathcal{C}}_{DM}$)

***Proposition 2:*** Let $\rho$ be any pure-state density matrix in $\mathcal{H}^d$ with its coherence vector $\boldsymbol{\mu}(\rho) = (\rho_{11}, \rho_{22}, ..., \rho_{dd})^T$, where $\{\rho_{ii}\}_{i=1}^{d}$ are the diagonal entries of $\rho$. Then, the following function $\check{\mathcal{C}}_{DM}(\rho)$ containing the multiplications between two different elements of $\boldsymbol{\mu}(\rho)$ in all possible combinations, acts as a coherence measure of $\rho$, i.e.,

$$\check{\mathcal{C}}_{DM}(\rho) = \sum_{i=1}^{d}\sum_{j=i+1}^{d}\rho_{ii}\rho_{jj} . \quad (4)$$

*Proof.* Like $\check{\mathcal{C}}_{DD}$, $\check{\mathcal{C}}_{DM}$ also needs to satisfy *conditions-1* to *condition-4*, to be accepted as a PSCM.

*Proof of condition-1.* Let $\rho \in \mathfrak{T}$; then $\boldsymbol{\mu}(\rho) \in \{P_\pi((1,0,...,0)^T)\}$. As $\boldsymbol{\mu}(\rho)$ possess only one non-zero element, all the $C_2^d$ number of $\rho_{ii}\rho_{jj}$ terms in the RHS of Eq. (4) give zero. Hence, $\check{\mathcal{C}}_{DM}(\rho) = 0$.

*Proof of condition-2.* As $\rho_{ii} \geq 0 \forall \rho_{ii}$ ($i \in \{1,2,...,d\}$) and the RHS of Eq. (4) gives the sum of all possible $\rho_{ii}\rho_{jj}$, it is not difficult to accept that $\check{\mathcal{C}}_{DM}(\boldsymbol{\mu}(\rho))$ is invariant under any permutation operation, i.e., $\check{\mathcal{C}}_{DM}\left(P_\pi(\boldsymbol{\mu}(\rho))\right) = \check{\mathcal{C}}_{DM}(\boldsymbol{\mu}(\rho)) \forall \pi \in \{1,2,...,d\}$.

*Proof of condition-3.* The domain of $\check{\mathcal{C}}_{DM}(\cdot)$ is the set of vectors $\boldsymbol{\mu}(\rho) \in \Omega$, which is a convex set (defined before); alongside, as $\check{\mathcal{C}}_{DM}(\boldsymbol{\mu}(\rho))$ comprises the sum of $\{\rho_{ii}\rho_{jj}\}$, it is *concave* by nature. For verification, two examples are the following: Let, $\boldsymbol{\mu}(\rho_1) = (0.2, 0.3, 0.5)^T$, $\boldsymbol{\mu}(\rho_2) = (0.4, 0.3, 0.3)^T$ and $\lambda = 0.5$; then $\check{\mathcal{C}}_{DM}(\lambda \boldsymbol{\mu}(\rho_1) + (1-\lambda)\boldsymbol{\mu}(\rho_2)) = 0.33$, whereas, $\lambda \check{\mathcal{C}}_{DM}(\boldsymbol{\mu}(\rho_1)) + (1-\lambda) \check{\mathcal{C}}_{DM}(\boldsymbol{\mu}(\rho_2)) = 0.32$, thus obeying *concavity*. In another example, let $\lambda = 0.25$ keeping the *coherence vectors* unchanged. Here again, the result follows concavity by fulfilling Eq. (3): $\check{\mathcal{C}}_{DM}(\lambda\boldsymbol{\mu}(\rho_1) + (1-\lambda)\boldsymbol{\mu}(\rho_2)) = 0.3325$ and $\lambda\,\check{\mathcal{C}}_{DM}(\boldsymbol{\mu}(\rho_1)) + (1-\lambda)\,\check{\mathcal{C}}_{DM}(\boldsymbol{\mu}(\rho_2)) = 0.3250$.

*Proof of condition-4.* A $d$-dimensional pure state $\rho$ is *MCS* only if all its diagonal elements are of the same value $\frac{1}{d}$; thus, $\boldsymbol{\mu}(MCS) = \left(\frac{1}{d},\frac{1}{d},\ldots,\frac{1}{d}\right)^T$. To prove that $\check{\mathcal{C}}_{DM}(MCS)$ is the maximal, it is first necessary to prove the following lemma:

*Lemma 1.* If the sum of a finite number of positive-valued variables is constant, then the sum of the product of two different variables in all possible ways is maximal if all the variables are of equal value.

*Proof of Lemma-1.* Let $(a,b,c)$ be a three-variable system with $a,b,c \geq 0$ and $a + b + c = S$, where $S$ is any positive constant (of finite value). Because of the summation condition, all three variables are bound within 0 and $S$, i.e., $0 \leq a \leq S$, $0 \leq b \leq S$ and $0 \leq c \leq S$. Therefore, it is easy to see that the multiplication of all these variables, i.e., $abc$ gives maximum only if $a = b = c = \frac{S}{3}$. Again, to make $ab$ maximum, the requirements are $a = b = \frac{S}{2}$ and $c = 0$; likewise, for $bc$ these are $a = 0$ and $b = c = \frac{S}{2}$. Now, let $M^{(3)} = ab + ac + bc = a(S-a) + bc$; for maximum $bc$, the first term ($a(S-a)$) will have no contribution as $a = 0$, thus $M^{(3)} = \frac{S}{4}$; whereas the first term is maximum when $a = \frac{S}{2}$ and $b + c = \frac{S}{2}$ (it can be simply achieved by taking the derivative of the first term over $a$ and then finding the maxima by equating it to zero). In this case, $M^{(3)} = \frac{S}{4} + \frac{S}{16} = \frac{5S}{16}$ with $a = \frac{S}{2}$, $b = c = \frac{S}{4}$. From these two instances, it can be inferred that $M$ is maximal when $\frac{S}{4} \leq a,b,c \leq \frac{S}{2}$. As all these three variables are evenly distributed in $M^{(3)}$ (it is easy to check that each of these three variables appears twice in $M^{(3)}$), the maximal is achieved only when $a = b = c = \frac{S}{3}$ (same as the maximal condition for $abc$) with $M_{max}^{(3)} = C_2^3 \frac{S^2}{9} = \frac{S^2}{3}$. Similar justification is applicable to any finite $n$-variable system where $M_{max}^{(n)} = C_2^n \frac{S^2}{n^2}$ with all the variables equal in magnitude, i.e., $\frac{S}{n}$. Hence, the proof of *lemma-1* is complete.

Going back to the pure state $\rho$, it is easy to see that the elements of $\boldsymbol{\mu}(\rho)$ are constrained by two conditions: $\sum_{i=1}^{d} \rho_{ii} = 1$ and $\rho_{ii} \geq 0\ \forall \rho_{ii}$ (as $\rho_{ii}$ is the probability of the basis state $|i\rangle$). Therefore, it is apparent from *lemma-1* that $\check{\mathcal{C}}_{DM}(\rho) = \sum_{i=1}^{d}\sum_{j=i+1}^{d} \rho_{ii}\rho_{jj}$ is maximal only when $\boldsymbol{\mu}(\rho) = \left(\frac{1}{d},\frac{1}{d},\ldots,\frac{1}{d}\right)^T$ and $\check{\mathcal{C}}_{DM}(MCS) = C_2^n \frac{1}{d^2}$.

Alternatively, *condition-4* can be directly proved with the help of the $l_1$-norm of coherence ($\mathcal{C}_{l_1}$). $\mathcal{C}_{l_1}$ is defined by the sum of the modulus of all the off-diagonal elements of $\rho$, i.e., $\mathcal{C}_{l_1}(\rho) = \sum_{i,j=1; i\neq j}^{d} |\rho_{ij}|$ [21]. Now, for any pure state, $|\rho_{ij}| = \sqrt{\rho_{ii}\rho_{jj}}$; therefore, $\mathcal{C}_{l_1}(\rho)$ can be rewritten as below:

$$\mathcal{C}_{l_1}(\rho) = 2\sum_{i=1}^{d}\sum_{j=i+1}^{d}\sqrt{\rho_{ii}\rho_{jj}}\,. \tag{5}$$

This expression of $\mathcal{C}_{l_1}$ is very similar to $\check{\mathcal{C}}_{DM}$ (Eq. (4)) except for the presence of twice multiplicity and the power-root of two (associated with each term in the summation) for $\mathcal{C}_{l_1}$. As $\mathcal{C}_{l_1}$ is a coherence measure, it obeys the criterion C5, i.e., $\mathcal{C}_{l_1}(\rho)$ is maximal only if $\rho \in MCS$; in other words, $\mathcal{C}_{l_1}$ obeys *condition-4*. Therefore, the close resemblance between the expressions of $\mathcal{C}_{l_1}(\rho)$ and $\check{\mathcal{C}}_{DM}(\rho)$ makes it obvious that $\check{\mathcal{C}}_{DM}(\rho)$ is maximal only if $\rho \in MCS$, i.e., when $\boldsymbol{\mu}(\rho) = \left(\frac{1}{d},\frac{1}{d},\ldots,\frac{1}{d}\right)^T$. With this, the proof of $\check{\mathcal{C}}_{DM}(\rho)$ being a valid PSCM is completed. ∎

### C. Coherence Monotones (or Measures) from $\check{\mathcal{C}}_{DD}$ and $\check{\mathcal{C}}_{DM}$

In the previous sub-section, we established $\check{\mathcal{C}}_{DD}$ and $\check{\mathcal{C}}_{DM}$ as two PSCMs. Therefore, both $\check{\mathcal{C}}_{DD}$ and $\check{\mathcal{C}}_{DM}$ are defined for any pure state. Now, in the more general context of mixed-state domain, we can safely apply these to the quantifier classes like $\mathcal{C}_P$ or $\mathcal{C}_{conv.roof}$, so that we achieve new *coherence monotones* or *measures* (depending on the natures of $\mathcal{C}_P$ and $\mathcal{C}_{conv.roof}$). In the following, we define two such new *coherence monotones* based on $\mathcal{C}_P$: diagonal difference of coherence ($C_P^{DD}$) and diagonal multiplication of coherence ($C_P^{DM}$).

a. *Diagonal difference of coherence ($C_P^{DD}$):*
*Definition 1.* If $\rho$ is a given state (pure or mixed) with its associated pure state set $R(\rho)$, then $C_P^{DD}(\rho)$ is a *coherence monotone* with

$$C_P^{DD}(\rho) = \inf_{|\psi\rangle \in R(\rho)} \check{C}_{DD}(|\psi\rangle)$$
$$= (d-1) - f(DD), \quad (6)$$

where $f(DD) = \sup_{|\psi\rangle \in R(\rho)} \sum_{i=1}^{d}\sum_{j=i+1}^{d}|\varphi_{ii} - \varphi_{jj}|$ with $\varphi_{ii(jj)} = |\psi\rangle\langle\psi|_{ii(jj)}$; Here, the minimal of all the $\check{C}_{DD}(|\psi\rangle \in R(\rho))$ is considered the measure of $\rho$.

b. *Diagonal multiplication of coherence ($C_P^{DM}$):*
*Definition 2.* If $\rho$ is a given state (pure or mixed) with its associated pure states set $R(\rho)$, then $C_P^{DM}(\rho)$ is a *coherence monotone* with

$$C_P^{DD}(\rho) = \inf_{|\psi\rangle \in R(\rho)} \check{C}_{DM}(|\psi\rangle)$$
$$= \inf_{|\psi\rangle \in R(\rho)} \sum_{i=1}^{d}\sum_{j=i+1}^{d} \varphi_{ii}\varphi_{jj}, \quad (7)$$

where $\varphi_{ii(jj)} = |\psi\rangle\langle\psi|_{ii(jj)}$.

Note that for any pure state $\rho = |\psi\rangle\langle\psi|$, both Eq. (6) and Eq. (7) reduce to the following forms, respectively: $C_P^{DD}(\rho) = \check{C}_{DD}(|\psi\rangle)$ and $C_P^{DM}(\rho) = \check{C}_{DM}(|\psi\rangle)$.

In a similar fashion, $\check{C}_{DD}$ and $\check{C}_{DM}$ provide two new *coherence measures* when applied in the *coherence measure class* $C_{conv.roof}$. The definitions of these two measures, $C_{conv.roof}^{DD}$ and $C_{conv.roof}^{DM}$, are presented below in a compact form below:

$$C_{conv.roof}^{j}(\rho) = \inf_{\{p_i,|\psi_i\rangle\}} \sum_i p_i \check{C}_j(|\psi_i\rangle), \quad (8)$$

with $j = 1 \, or \, 2$ stands for "$DD$" and "$DM$," respectively. Here, the infimum is taken over all the pure state decompositions of $\rho$ ($\rho = \sum_i p_i(|\psi_i\rangle)$).

Note that we do not consider $C_P^{DD}$ or $C_P^{DM}$ as a *coherence measure* but rather classify them as *coherence monotones*. The reason will be clarified in the next section.

## IV. THE CONVEXITY

According to (C4), $C(\cdot)$ is convex under mixing iff $C(\rho) \leq \sum p_i C(\rho_i)$ with $\rho = \sum p_i \rho_i$, where $\rho_i$ are the participating pure states with their mixing probabilities $p_i$. It is not difficult to see that, by definition, $C_{conv.roof}$ supports convexity (C4) as $C_{conv.roof}(\rho) = \inf_{\{p_j,|\rho_j\rangle\}} \sum_j p_j \check{C}(\rho_j) \leq \sum p_i \check{C}(\rho_i)$, irrespective of any valid $\check{C}$ and for any $\rho$. Therefore, $C_{conv.roof}^{DD(DM)}$ naturally falls under the *coherence measure* category, obeying all five criteria (C1) – (C5).

However, in the case of $C_P$, it has been shown (see *Theorem 3* of [32]) that $C_P$ is convex if, for any ensemble $\{p_i, \rho_i\}$ of $\rho$, there always exists a pure state $|\psi_0\rangle \in R(\rho)$ such that $\check{C}(|\psi_0\rangle) \leq \sum p_i \check{C}(\rho_i)$. Putting it another way, $C_P$ is convex if $C_P(\rho) = C_{conv.roof}(\rho) \, \forall \rho$ (*Theorem 4* of ref. [32]). Here, through a few logical steps, we are going to explain why this condition cannot be met for any bona fide $\check{C}$; in other words, why $C_P$ is not convex in general. But before that, we redefine $C_P$, so that the remaining task becomes easier.

The *theorem 2* of ref. [32] shows that the set of pure states $|\psi\rangle \in R(\rho)$ (see *Theorem 1*) corresponding to $\rho$, can effectively be narrowed down to the subset $Q(\rho)$ that also contains the optimal pure state $|\psi_o\rangle$, i.e., $|\psi_o\rangle \in Q(\rho) \subset R(\rho)$. In the following, we show that a subset $M(\rho) \subseteq Q(\rho)$ is also available that contains $|\psi_o\rangle$. Before coming to that, we briefly recap the theory of *majorization* as an important prerequisite.

***Majorization*:** A vector $\boldsymbol{p} \in \mathbb{R}^n$ majorizes [41, 42] another vector $\boldsymbol{q} \in \mathbb{R}^n$ or $\boldsymbol{p} \succcurlyeq \boldsymbol{q}$ if

$$\sum_{i=1}^{l} p_i^{\downarrow} \geq \sum_{i=1}^{l} q_i^{\downarrow} \; \forall l \in \{1,2,\ldots,n\}, \quad (9)$$

where $p_i^\downarrow$ and $q_i^\downarrow$ are the $i$-th elements of $\boldsymbol{p}$ and $\boldsymbol{q}$, respectively, when both rearranged in descending orders, i.e., $p_1^\downarrow \geq p_2^\downarrow \geq \cdots \geq p_n^\downarrow$, and $q_1^\downarrow \geq q_2^\downarrow \geq \cdots \geq q_n^\downarrow$.

In quantum resource theory (QRT), *majorization* is being used as an important though simple mathematical tool that determines the transformation from one quantum state $\varphi \in \mathcal{H}^d$ to another state $\rho \in \mathcal{H}^d$ via free or incoherent operations (IO); if $\varphi \xrightarrow{IO} \rho$, then it must be that $\boldsymbol{\mu}(\rho) \succcurlyeq \boldsymbol{\mu}(\varphi)$ [20, 25, 40]. Again, according to the postulates of the resource theory of coherence, if $\varphi \xrightarrow{IO} \rho$, then it is obvious that $\mathcal{C}(\varphi) \geq \mathcal{C}(\rho)$ [21]. Therefore, we can translate this very concept of free operation into majorization through the following statement:

$$\varphi \xrightarrow{IO} \rho \text{ iff } \mu(\rho) \succcurlyeq \mu(\varphi) \Rightarrow \sum_{i=1}^l \mu^\downarrow(\rho) \geq \sum_{i=1}^l \mu^\downarrow(\varphi) \; \forall l \in \{1, \ldots, d\}. \quad (10)$$

With this piece of information in hand, we are now ready to discuss the further redefinition of $\mathcal{C}_P$.

**Theorem 2.** The coherence monotone $\mathcal{C}_P(\rho)$ of a density matrix $\rho$ with its pure state decompositions $\{p_i, |\psi_i\rangle\}$ can be redefined as

$$\mathcal{C}_P(\rho) = \underset{|\varphi\rangle \in M(\rho)}{all} \check{\mathcal{C}}(|\varphi\rangle), \quad (11)$$

where $\check{\mathcal{C}}(\cdot)$ is any bona fide PSCM, and $M(\rho)$ is the set of pure states $|\varphi\rangle$ that fulfill the following relation:

$$\boldsymbol{\mu}^\downarrow(|\varphi\rangle) = \underset{\{p_i, |\psi_i\rangle\}}{\sup} \sum_i p_i \boldsymbol{\mu}^\downarrow(|\psi_i\rangle), \quad (12)$$

where the supremum is considered based on *majorization*.

*Proof of Theorem 2.* Let there exist 3 possible pure-state decompositions of $\rho$: $\sum_i p_i(|\psi_i\rangle), \sum_j p_j(|\psi_j\rangle), \sum_k p_k(|\psi_k\rangle)$; again, let $\sum_i p_i \boldsymbol{\mu}^\downarrow(|\psi_i\rangle) \preccurlyeq \sum_j p_j \boldsymbol{\mu}^\downarrow(|\psi_j\rangle) \preccurlyeq \sum_k p_k \boldsymbol{\mu}^\downarrow(|\psi_k\rangle)$ (without loss of generality). Now, *theorem 2* of ref. [32], says that the pure states $|\vartheta\rangle$ corresponding to all the different decompositions of $\rho$ collectively form the pure-state set $Q(\rho)$ that contains the optimal state $|\psi_o\rangle$: $|\vartheta\rangle \in Q(\rho) \subset R(\rho)$, and for any member $|\vartheta_s\rangle$ of $Q(\rho)$ there exists a decomposition $\{p_s, |\psi_s\rangle\}$ of $\rho$ such that $\boldsymbol{\mu}^\downarrow(|\vartheta_s\rangle) = \sum_i p_s \boldsymbol{\mu}^\downarrow(|\psi_s\rangle)$. It implies that each of the different decompositions of $\rho$ corresponds to a pure-state set. In this case, let $\{|\varphi^m\rangle\}$ (with $m \in \{i, j, k\}$) represent three such sets for the three decompositions, where for any element $|\varphi_l^m\rangle$ in the set $\{|\varphi^m\rangle\}$, the following relation holds:

$$\boldsymbol{\mu}^\downarrow(|\varphi_l^m\rangle) = \sum_m p_m \boldsymbol{\mu}^\downarrow(|\psi_m\rangle) \; \forall |\varphi_l^m\rangle \in \{|\varphi^m\rangle\}. \quad (13)$$

Thus, $\{|\varphi^m\rangle\} \subseteq Q(\rho)$ for all $m$; and $\{|\varphi^i\rangle\} \cup \{|\varphi^j\rangle\} \cup \{|\varphi^k\rangle\} = Q(\rho)$. Again, from the connection between the *majorization* relation and incoherent transformation (see Eq. (10)), it is clear that $\{|\varphi^i\rangle\} \xrightarrow{IO} \{|\varphi^j\rangle\} \xrightarrow{IO} \{|\varphi^k\rangle\} \xrightarrow{IO} \sum_k p_k(|\psi_k\rangle)$. As $\mathcal{C}_P$ always goes for the minimal coherent state, the subset $\{|\varphi^k\rangle\} = M(\rho)$ of $Q(\rho)$ only holds the optimum state $|\psi_o\rangle$; thus, $|\psi_o\rangle \in M(\rho) \subseteq Q(\rho) \subset R(\rho)$.

Next, to get the optimum state $|\psi_o\rangle$ from $M(\rho)$, it can be emphasized that the coherence vectors in descending order ($\boldsymbol{\mu}^\downarrow$) for all the elements $|\varphi_l^m\rangle$ in the $m$-th subset $\{|\varphi^m\rangle\}$ are identical (Eq. (13)). It means that all the pure states in $M(\rho)$ are of the same coherence; the coherence vectors of any two states in $M(\rho)$ only differ by a permutation operation $P_\pi$. It means that all the states in $M(\rho)$ can produce the optimum result; therefore, Eq. (11) is true. The same logic can be applied to any number of possible pure-state decompositions of $\rho$, hence proved. ∎

**Theorem 3.** In general, $\mathcal{C}_P$ does not obey convexity (the criterion (C4)). Hence, it cannot be treated as a *coherence-measure* class.

*Proof of Theorem 3.* It is seen from *Theorem 2* that $\mathcal{C}_P(\rho) = \check{\mathcal{C}}(|\varphi\rangle)$ where $|\varphi\rangle$ is any pure state that belongs to $M(\rho)$ ($|\varphi\rangle \in M(\rho)$). $|\varphi\rangle$ is related to the pure-state decomposition of $\rho$ through the coherence vector s.t. $\boldsymbol{\mu}^\downarrow(|\varphi\rangle) = \underset{\{p_i, |\psi_i\rangle\}}{\sup} \sum_i p_i \boldsymbol{\mu}^\downarrow(|\psi_i\rangle)$. According to the *majorization* principle (see Eqs. (9)

and (10)), the coherence vector for the decomposition that majorizes all other decompositions of $\rho$, gives the least coherence. Thus, $|\varphi\rangle$ points to the optimal decomposition for $\mathcal{C}_P$, let say $\{p_k, |\psi_k\rangle\}$, $\mathcal{C}_P(\sum_k p_k(|\psi_k\rangle)) = \mathcal{C}_P(\rho) = \check{\mathcal{C}}(|\varphi\rangle)$. As a result,

$$\boldsymbol{\mu}^\downarrow(|\varphi\rangle) = \sum_k p_k \boldsymbol{\mu}^\downarrow(|\psi_k\rangle)$$

Thus,

$$\check{\mathcal{C}}(\boldsymbol{\mu}^\downarrow(|\varphi\rangle)) = \check{\mathcal{C}}(\sum_k p_k \boldsymbol{\mu}^\downarrow(|\psi_k\rangle)$$

$$\geq \sum_k p_k \check{\mathcal{C}}(\boldsymbol{\mu}^\downarrow(|\psi_k\rangle)). \quad (14)$$

The inequality in Eq. (14) comes from the *concavity* (*condition-3*) of any valid PSCM, $\check{\mathcal{C}}$. As $\check{\mathcal{C}}$ must be invariant under any permutation (*condition-2*), $\check{\mathcal{C}}(\boldsymbol{\mu}^\downarrow(|\theta\rangle)) = \check{\mathcal{C}}(\boldsymbol{\mu}|\theta\rangle)$ for any pure state $|\theta\rangle$, then from Eqs. (11) and (14)

$$\check{\mathcal{C}}(\boldsymbol{\mu}^\downarrow(|\varphi\rangle)) = \check{\mathcal{C}}(|\varphi\rangle) = \mathcal{C}_P(\sum_k p_k(|\psi_k\rangle))$$

$$\geq \sum_k p_k \check{\mathcal{C}}(|\psi_k\rangle). \quad (15)$$

Eq. (15) implies that $\mathcal{C}_P$ does not satisfy convexity (C4) in general. As $\mathcal{C}_P$ fulfills all the remaining criteria C1-C3 and C5 [32], it is a bona fide *coherence monotone class*. However, it cannot be treated as a *coherence-measure class.* ∎

Based on *Theorem 3,* one can remark that $\mathcal{C}_P^{DD}$ and $\mathcal{C}_P^{DM}$ are two new *coherence monotones* (not *coherence measures*). However, there is no problem with considering $\mathcal{C}_{conv.roof}^{DD}$ and $\mathcal{C}_{conv.roof}^{DM}$ as two new *coherence measures*.

## V. DISCUSSION AND CONCLUSIONS

The pure-state coherence measures (PSCMs) are the key elements for the coherence quantification classes, like quantification via pure-state coherence ($\mathcal{C}_P$) or the convex roof technique ($\mathcal{C}_{conv.roof}$). Based on these techniques, PSCMs endow us with different coherence monotones or measures, thus contributing towards the enrichment of the resource theory of coherence. In this work, we have presented two such novel PSCMs and proved their validity through the fulfillment of all four necessary conditions (given in Sec. III). As the name suggests, *the diagonal difference of PSCM* ($\check{\mathcal{C}}_{DD}$, see *Proposition 1*) depends on the modulus of the difference between any two elements (Eq. (2)) of the coherence vector $\boldsymbol{\mu}$ (which contains all the diagonal elements of a density matrix), whereas in *diagonal multiplication of PSCM* ($\check{\mathcal{C}}_{DM}$, see *Proposition 2*), the main ingredients are the possible factors between two different elements of $\boldsymbol{\mu}$ (Eq. (4)). Moreover, we have presented the forms $\check{\mathcal{C}}_{DD}$ and $\check{\mathcal{C}}_{DM}$ achieve (Eqs. (6), (7), and (8)) as *coherence monotones* (or *measures*) when incorporated into $\mathcal{C}_P$ and $\mathcal{C}_{conv.roof}$. Note that the usage of $\check{\mathcal{C}}_{DD}$ and $\check{\mathcal{C}}_{DM}$ is not limited to $\mathcal{C}_P$ or $\mathcal{C}_{conv.roof}$; these are equally applicable to any other bona fide coherence measure (monotone) classes.

Quantification via pure-state coherence ($\mathcal{C}_P$) offers a unique method that considers the least coherent state from the pure-state set $R(\rho)$ or its subset $Q(\rho)$ [32] to quantify $\rho$. We have further redefined $\mathcal{C}_P$ by showing that $M(\rho)$, a subset of $Q(\rho)$, is also available that contains the required pure state of minimal coherence (*Theorem 2*). The pure states $|\varphi\rangle \in M(\rho)$ are connected to the minimal decomposition (in terms of coherence) of $\rho$ (let say $\{p_k, |\psi_k\rangle\}$) through majorization, s.t. $\boldsymbol{\mu}^\downarrow(|\varphi\rangle) = \sum_k p_k \boldsymbol{\mu}^\downarrow(|\psi_k\rangle)$. In that sense, the concept of $\mathcal{C}_P$ is quite similar to the traditional technique of $\mathcal{C}_{conv.roof}$; both rely on $\{p_k, |\psi_k\rangle\}$. Regarding convexity, we have shown that $\mathcal{C}_P$ cannot fulfill the criterion C4 due to the *concave* nature of any valid PSCM; therefore, $\mathcal{C}_P$ is not a *coherence-measure class* in general.

We know that a bona fide *coherence measure* can be reduced to a PSCM (a symmetric, concave function that fulfills all the necessary criteria, from *condition-1* to *condition-4*) [39, 40]. In other words, a valid PSCM may have its *coherence-measure* counterpart. Therefore, with $\check{\mathcal{C}}_{DD}$ and $\check{\mathcal{C}}_{DM}$ in hand, the search for their *coherence-measure* equivalents with respective operational and methodological aspects for completeness can be a topic of further research.

### ACKNOWLEDGEMENT

We thank S. Goswami for language correction and editing. DG acknowledges funding support from MEITY, SERB, and STC ISRO of the Govt. of India.

[1] M.-L. Hu, X.-M. Wang, and H. Fan, Phys. Rev. A 98, 032317 (2018).

[2] J. Ma, B. Yadin, D. Girolami, V. Vedral, and M. Gu, Phys. Rev. Lett. 116, 160407 (2016).

[3] A. Streltsov, U. Singh, H. S. Dhar, M. N. Bera, and G. Adesso, Phys. Rev. Lett. 115, 020403 (2015).

[4] E. Chitambar and M.-H. Hsieh, Phys. Rev. Lett. 117, 020402 (2016).

[5] H. Zhu, M. Hayashi, and L. Chen, Phys. Rev. A 97, 022342 (2018).

[6] I. Marvian and R. W. Spekkens, Nat. Commun. 5, 3821 (2014).

[7] M. Lostaglio, K. Korzekwa, D. Jennings, and T. Rudolph, Phys. Rev. X 5, 021001 (2015).

[8] G. Gour, M. P. Müller, V. Narasimhachar, R. W. Spekkens, and N. Y. Halpern, Phys. Rep. 583, 1 (2015).

[9] L. Rybak, S. Amaran, L. Levin, M. Tomza, R. Moszynski, R. Kosloff, C. P. Koch, and Z. Amitay, Phys. Rev. Lett. 107, 273001 (2011).

[10] A. Misra, U. Singh, S. Bhattacharya, and A. K. Pati, Phys. Rev. A 93, 052335 (2016).

[11] K. Brandner, M. Bauer, and U. Seifert, Phys. Rev. Lett. 119, 170602 (2017).

[12] Z. Wang, W. Wu, G. Cui, and J. Wang, New J. Phys. 20, 033034 (2018).

[13] C. Zhang, B. Yadin, Z.-B. Hou, H. Cao, B.-H. Liu, Y.-F. Huang, R. Maity, V. Vedral, C.-F. Li, G.-C. Guo, and D. Girolami, Phys. Rev. A 96, 042327 (2017).

[14] P. Giorda and M. Allegra, J. Phys. A: Math. Theor. 51, 025302 (2017).

[15] B. Huttner, N. Imoto, N. Gisin, and T. Mor, Phys. Rev. A 51, 1863 (1995).

[16] J. Ma, Y. Zhou, X. Yuan, and X. Ma, Phys. Rev. A 99, 062325 (2019).

[17] E. M. Gauger, E. Rieper, J. J. L. Morton, S. C. Benjamin, and V. Vedral, Phys. Rev. Lett. 106, 040503 (2011).

[18] C.-M. Li, N. Lambert, Y.-N. Chen, G.-Y. Chen, and F. Nori, Sci. Rep. 2, 885 (2012).

[19] A. Streltsov, G. Adesso, and M. B. Plenio, Rev. Mod. Phys. 89, 041003 (2017).

[20] E. Chitambar and G. Gour, Rev. Mod. Phys. 91, 025001 (2019).

[21] T. Baumgratz, M. Cramer, and M. B. Plenio, Phys. Rev. Lett. 113, 140401 (2014).

[22] X.-D. Yu, D.-J. Zhang, G. F. Xu, and D. M. Tong, Phys. Rev. A 94, 060302(R) (2016).

[23] M.-L. Hu, X. Hu, J. Wang, Y. Peng, Y.-R. Zhang, and H. Fan, Phys. Rep. 762–764, 1 (2018).

[24] D. K. L. Oi and J. Åberg, Phys. Rev. Lett. 97, 220404 (2006).

[25] A. Winter and D. Yang, Phys. Rev. Lett. 116, 120404 (2016).

[26] A. E. Rastegin, Phys. Rev. A 93, 032136 (2016).

[27] C.-S. Yu, Phys. Rev. A 95, 042337 (2017).

[28] S. Rana, P. Parashar, and M. Lewenstein, Phys. Rev. A 93, 012110 (2016).

[29] Y. Yao, X. Xiao, L. Ge, and C. P. Sun, Phys. Rev. A 92, 022112 (2015).

[30] L.-H. Shao, Z. Xi, H. Fan, and Y. Li, Phys. Rev. A 91, 042120 (2015).

[31] Yuan, X., H. Zhou, Z. Cao, and X. Ma, Phys. Rev. A 92, 022124 (2015).

[32] Deng-hui Yu, Li-qiang Zhang, and Chang-shui Yu, Phys. Rev. A 101, 062114 (2020).

[33] Z.-W. Liu, X. Hu, and S. Lloyd, Resource Destroying Maps, Phys. Rev. Lett. 118, 060502 (2017).

[34] K. Bu, U. Singh, S.-M. Fei, A. K. Pati, and J. Wu, Phys. Rev. Lett. 119, 150405 (2017).

[35] Q. Zhao, Y. Liu, X. Yuan, E. Chitambar, and X. Ma, Phys. Rev. Lett. 120, 070403 (2018).

[36] Z.-W. Liu, K. Bu, and R. Takagi, Phys. Rev. Lett. 123, 020401 (2019).

[37] Y. Peng, Y. Jiang, and H. Fan, Phys. Rev. A 93, 032326 (2016).

[38] M. B. Plenio, Phys. Rev. Lett. 95, 090503 (2005).

[39] S. Du, Z. Bai, and X. Qi, Quantum Inf. Comput. 15, 1307 (2015).

[40] H. Zhu, Z. Ma, Z. Cao, S.-M. Fei, and V. Vedral, Phys. Rev. A 96, 032316 (2017).

[41] R. Bhatia, Matrix Analysis ,(Springer, New York ,1996).

[42] Marshall, A. W., I. Olkin, and B. Arnold, , Inequalities: Theory of Majorization and Its Applications (Springer, New York, 2011).